# Island hopping of active colloids[†]

Venkata Manikantha Sai Ganesh Tanuku[1]*, Peter Vogel[1], Thomas Palberg[1] and Ivo Buttinoni[2]

[1] Institute of Physics, Johannes Gutenberg University, 55128 Mainz, Germany
[2] Institute for Experimental Physics of Condensed Matter, Heinrich-Heine University, 40225 Düsseldorf, Germany

* corresponding author e-mail: vtanuku@uni-mainz.de

**Abstract:** Individual self-propelled colloidal particles, like active Brownian particles (ABP) or run-and-tumble swimmers (RT), exhibit characteristic and well-known motion patterns. However, their interaction with obstacles remains an open and important problem. We here investigate the two-dimensional motion of silica-gold Janus particles (JP) actuated by AC electric fields and suspended and cruising through silica particles organized in rafts by mutual phoretic attraction. A typical island contains dozens of particles. The JP travels straight in obstacle-free regions and reorients systematically upon approaching an island. As underlying mechanism, we tentatively propose a hydrodynamic torque exerted by the solvent flow towards the islands on the JP's local flow field, leading to an alignment of respective solvent flow directions. This systematic behavior is in contrast with the reorientation observed for free active Brownian particles and run-and-tumble microswimmers.

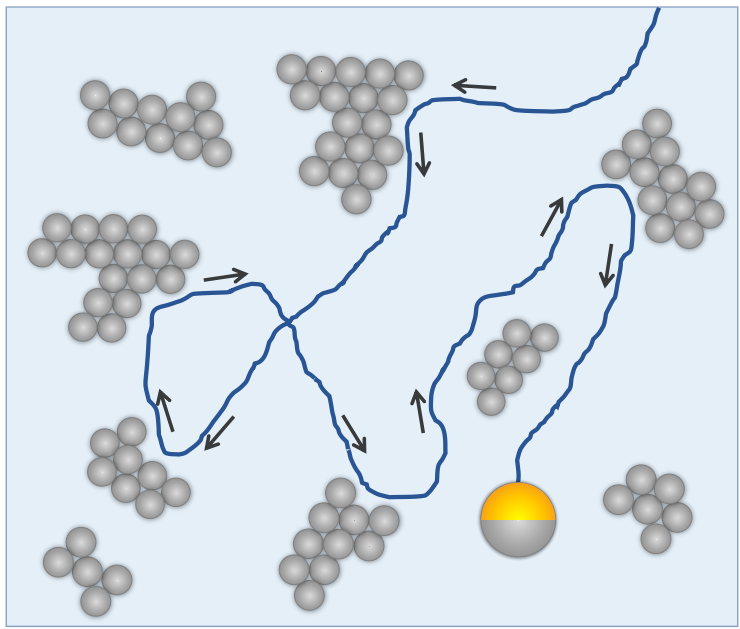

# 1 Introduction

The motility of active Brownian particles (ABP), i.e. synthetic microswimmers that undergo directed motion at the expense of consumed energy, has been a focal point of current research in the field of active matter.[1,2] In many experimental realizations, microspheres whose surface is made of two distinct materials, known as Janus particles (JP), sink to the bottom of a sample cell and self-propel in the *xy*-plane due to slip flows triggered by self-generated chemical or thermal gradients[3], or electric fields applied in the vertical direction (*z*).[4] In Newtonian fluids and in the absence of confinements or forces, the two-dimensional active Brownian motion proceeds at a velocity which is constant in magnitude, but whose direction is randomized by rotational diffusion.[2,5] The resulting trajectory is directed at short times but becomes diffusive for long times and long distances travelled. On the other hand, biological run-and-tumble swimmers (e.g., *E. coli*) show straight motion between discrete tumbling events, during which the swimmers reorient randomly. So, one observes again a direction persistence at short times, followed by diffusive behavior at longer time scales.[6]

Obstacles significantly alter the features of this motion. The short-time directional persistence leads to the sliding of self-propelling colloids along flat walls.[7,8] Further, by hydrodynamic interaction with spherical objects, such as 'passive' beads, the swimmers may orbit around the latter[9] or even bind to target particles or surfaces.[10] Upon meeting an extended obstacle head on, the directional persistence also suppresses an immediate escape: particles will accumulate until, for each, rotational diffusion reorients the direction of the velocity vector.[11,12] This has been exploited to power microfabricated devices[13], rectify the flow of colloidal suspensions[14], or exert pressure on boundaries.[15] For many accumulated ABPs, these themselves may act as obstacles. An active particle joining a cluster cannot leave within timescales that are much smaller than its characteristic rotational diffusion time ($\tau_R = (8\pi\eta R^3)/(k_B T)$, for microbeads of radius R and thermal energy $k_B T$, in a solvent of viscosity $\eta$). This underlies motility-induced phase transitions in active suspensions and segregation in mixtures of active and passive colloids.[16,17]

Thus far, the focus of research has laid on random reorientations. In the present work, we use JP showing directed motion in obstacle-free environments. We let them propel through a landscape of randomly distributed islands made of silica particles. JP motion and island formation are both triggered by the very same phenomenon: induced charge electroosmosis (ICEO) upon AC electric fields applied normal to the substrate surface. Once the silica beads are assembled into islands, a persistence solvent flow towards the island surface results, leading to a new type of microswimmer-obstacle interaction. We observe that, upon approach, the JPs reorient in a systematic way, such that their propulsion direction is turned from towards the obstacle to away from the obstacle. As the velocity magnitude hardly changes during turning, this process is loosely reminiscent of elastic scattering. JP motion is then continued in persistent direction until the encounter with the next island. The observed type of cruising shows characteristic differences as compared to the well-known ABP motion.

## 2 Materials and Methods

### 2.1 Colloidal particles

As 'passive' particles, we use commercial silica microspheres (SiO$_2$, radius R$_P$ = 1.46 µm, Microparticles GmbH). Janus particles (JPs) were synthesized using drop-casting method as reported by J. Yan et al.[18] First, the surface of a glass slide was treated with piranha solution (conc. H$_2$SO$_4$:H$_2$O$_2$: 1:3) for 3-4 hours, sonicated and rinsed with distilled water. Then, a dilute suspension of silica particles (SiO$_2$, radius R$_{JP}$ = 2.5 µm, Bangs Laboratories, USA) was spread over an inclined (45°) glass slide such that the solution flows down the slide leaving a monolayer of particles. After drying, we deposited onto the particles a thin layer of chromium (approx. 3nm) followed by a second thin layer of gold (approx. 15nm) using a thermal evaporator (Edwards BOC/Auto 306). Due to self-shadowing effects, only the top halves of the particles are coated. The JPs are finally detached from the glass substrate by sonicating the glass slide for a few seconds.

### 2.2 Sample preparation and imaging

The colloidal suspension is injected in a sample cell made of two co-planar transparent indium tin oxide coated glasses (ITO, negatively charged, surface resistivity 25 − 30 Ω/sq, Solems S. A.) separated by a circular spacer (Grace Bio-Labs secure seal of 120 µm height). Care was taken to avoid any residual air bubbles. The sample cell is then placed onto the stage of an inverted microscope (Leica DMI3000 B, Leica Microsystems GmbH, Wetzlar, Germany). The particles sink to the bottom ITO substrate and the system is allowed to equilibrate before imaging is performed in transmission. The top and bottom electrodes are connected to a function generator (Voltcraft MXG 9810) in order to apply an AC electric field with a fixed frequency $\mathbf{f = 1.5\ kHz}$ and peak-to-peak voltage $\mathbf{V_{pp}}$ between 0 and 10 V (electric field E between 0 and 0.083 V/µm). The motion of the active particles is recorded using a CMOS camera (Basler ACE) at 10-30 frames per second. A Python particle tracking code is finally used ex situ to find the particle positions and reconstruct the trajectories.

## 3 Experimental Results

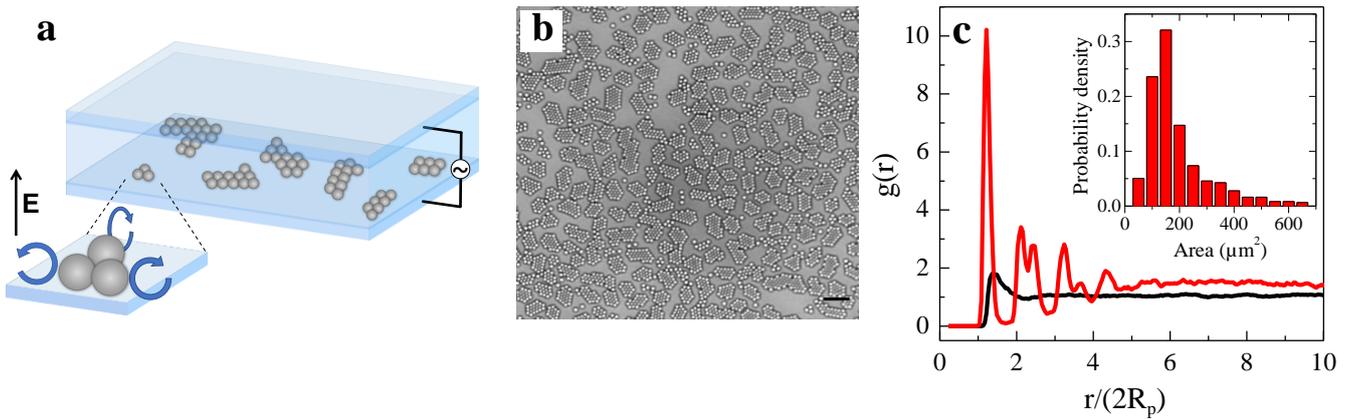

**Figure 1. Assembly of 'passive' silica microspheres under AC electric fields. (a)** Sketch of the experimental setup and EHD flows leading to mutual attractions between silica microspheres. **(b)** Microscopic image of silica passive particles ($R_P$ = 1.46 μm) forming close packed 'islands' at area fraction 0.42 in aqueous environments by applying an AC electric field in the vertical direction. The field switching frequency is kept constant at f = 1.5 kHz. The scale bar corresponds to 20 μm. **(c)** Radial distribution function g(r) plotted as a function of the normalized interparticle distance for E = 0 (black solid line) and E = 0.083 V/μm (red solid line). Inset: Probability density distribution of the area of the islands for E = 0.083 V/μm. The average cluster area is (426±80) μm².

We first characterize the 'passive environment' which consists in several large clusters of passive silica spheres (islands, see for instance **Fig. 1a** and **Fig. 1b**) forming in the xy-plane just above the bottom electrode when AC electric fields of frequency $\mathbf{f = 1.5\ kHz}$ are applied in the z-direction. The area fraction, $\mathbf{\phi = 0.42}$, is kept constant in all experiments. Prior to the application of electric field, we inject into the sample cell a fixed number of SiO$_2$ particles, which then sediment onto the bottom electrode and self-organize into a fluid phase with no translational or orientational order. The first peak of the radial distribution function g(r) is located at relative distances $\mathbf{r > 2R_p}$ (**Fig. 1c**, black data) indicating that, in the absence of electric fields, the pair interaction between passive SiO$_2$ colloids is repulsive. When we apply an AC electric field of amplitude greater than E = 0.016 V/μm, large crystalline close packed clusters (islands), as those shown in **Fig. 1b**, form. g(r) now exhibits sharp regular peaks due to the local hexagonal order. Its first maximum is shifted to $\mathbf{r \approx 2R_p}$ because the attraction triggered by the vertical electric field overcomes the electrostatic repulsion (**Fig. 1c**, red data). The formation of these islands only takes few seconds after the field is turned on (**Supplementary Video S1**). They are then mostly stable for more than 30 minutes, although few small clusters may merge into big clusters and others may rotate at slow speed. For larger times, the clusters eventually merge. The average cluster area and average distance between the clusters, measured using 10 different images, are $A_c$ = 426 ± 80 μm² and $d_c$ = 18 ± 7 μm, respectively (see Inset of **Fig. 1c**, where we plot the probability distribution of the area occupied by the individual clusters). Importantly, in the experimental range of electric fields (between 0.016 V/μm and 0.083 V/μm), the cluster size does not depend on the voltage and the spheres remain close to the bottom electrode. Our observations show that the clustering is driven by strong and long-range attractive EHD flows (see sketch in **Fig. 1a**) triggered by the electric field.[19] If the field is switched off, the islands 'dissolve' and the fluid phase is slowly recovered by thermal diffusion.

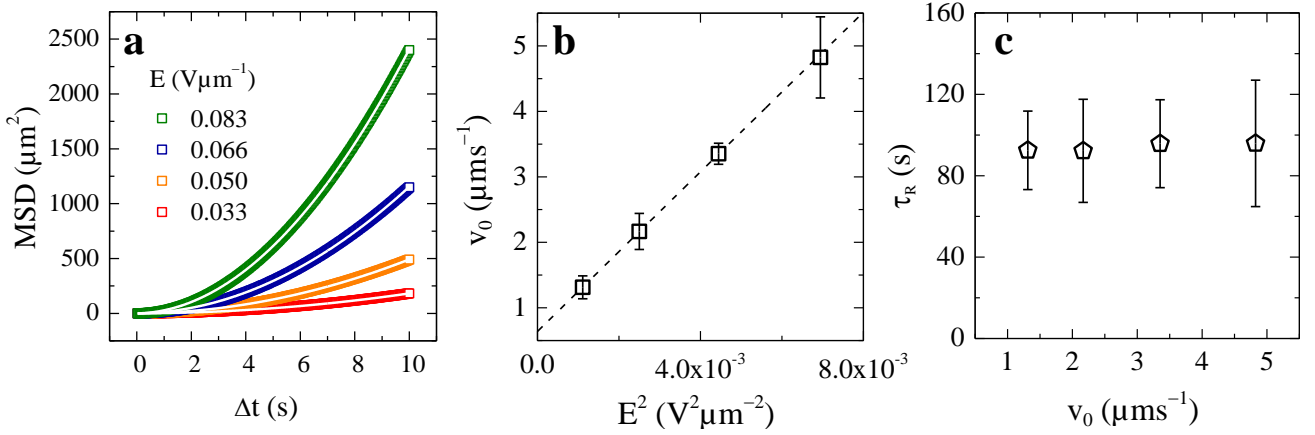

**Figure 2. Two-dimensional ABM of silica-gold JP without obstacles. (a)** Mean squared displacement (MSD) of JPs for increasing values of electric field (indicated in the legend) and frequency f = 1.5 kHz. The solid lines are least-squares fits of the data based on Equation (1). **(b)** Corresponding particle velocity, $v_0$, plotted against the squared electric-field strength. The dashed line is the least-squares linear fit of the data. **(c)** Characteristic reorientation time, $\tau_R$, of JPs moving in 2D as a function of $v_0$. Data in (a) – (c) are an average over 9 – 10 individual particles.

Silica-gold JPs also sink onto the bottom electrode and respond to the applied electric field. In the absence of passive obstacles, they undergo a free two-dimensional active Brownian motion, i.e., they swim at constant velocity $v_0$ with the $SiO_2$ hemisphere heading and reorient in 2D with a characteristic timescale $\tau_R$. The free-swimming velocity, $v_0$, is extracted by measuring the short-time mean squared displacement (MSD, **Figure 2a**) and fitting it according to:

$$\text{MSD}(\Delta t) = 4D\Delta t + v_0^2 \Delta t^2, \tag{1}$$

where $\Delta t$ is the delay time and D is the translational diffusion coefficient of the particles assumed to be 0.089 µm$^2$/s according to the Stokes-Einstein relation. As reported in previous works[18,20,21] and shown in **Figure 2b**, $v_0$ scales linearly with the squared amplitude of the applied AC electric field ($E^2$). Conversely, the reorientation time, $\tau_R$, is deduced from the full MSD of active Brownian particles, i.e.

$$\text{MSD}(\Delta t) = 4D\Delta t + 2v_0^2 \tau_R^2 \left(\frac{\Delta t}{\tau_R} - 1 + e^{-\Delta t/\tau_R}\right), \tag{2}$$

and is roughly independent of $v_0$ (**Figure 2c**). We obtain swimming velocities between 1 and 5 µm/s, which correspond to persistence lengths ($L = v_0 \cdot \tau_R$ – the typical length of straight paths) between 100 and 500 µm which is much larger than the distance between the clusters ($d_c$).

**Figure 3a** shows the trajectory of an active JP when it is surrounded by islands of silica colloids (**Figure 1**) and an AC electric field E = 0.066 V/µm is applied in z. The linear increase of the mean swimming velocity (v, Figure 3b) with $E^2$ is preserved, but v ~ 0.5$v_0$, indicating that, on average, the complex environment reduces the motility of the active colloids. By calculating the instantaneous velocity as a function of time for the trajectory shown in **Figure 3a**, we observed no significant change when they approach and leave the islands (**Supplementary Figure S1**). Nonetheless, the key difference with free active Brownian motion lies in the fact that, here, the direction of the velocity is randomized not only by rotational diffusion but also by microswimmer-island 'collisions'. In particular, since $\tau_R$ is much larger than the typical time required for the microswimmers to travel between islands (given by $d_c/v$) for all values of v used in our experiments, the contribution of rotational diffusion is negligible. Hence, the 2D trajectory of the active Brownian particle is characterized by the combination of straight paths and abrupt reorientations, and qualitatively resembles the run-and-tumble motion of bacteria such as *E. coli*. In **Fig. 3c**, we plot the angle θ of the instantaneous velocities of the trajectory in **Fig. 3a**. 'Runs' can be identified by the regions where θ is roughly constant. On the other hand, 'turnings' give rise to abrupt jumps lasting only few seconds (see colored circles). In several instances the changes of local angle are also greater than 90°. These observations are inconsistent with the 'classical' interaction of active colloids with clusters (leading for instance to motility-induced phase separation), according to which the microswimmers needs to wait a time $\sim \tau_R$ to break free from a cluster. Here, a close look at the JP-island interaction (**Fig. 3d–f**) reveals that, when a JP collides against one of the clusters, its orientation is swiftly reversed so that it leaves as if it were 'elastically' scattered by the island. It always swims in the direction given by the arrow linking the poles of the Au-hemisphere to the pole of the $SiO_2$-hemispheres, but tumbling events are predominantly driven by local flow fields rather than rotational diffusion (see **Discussion** below).

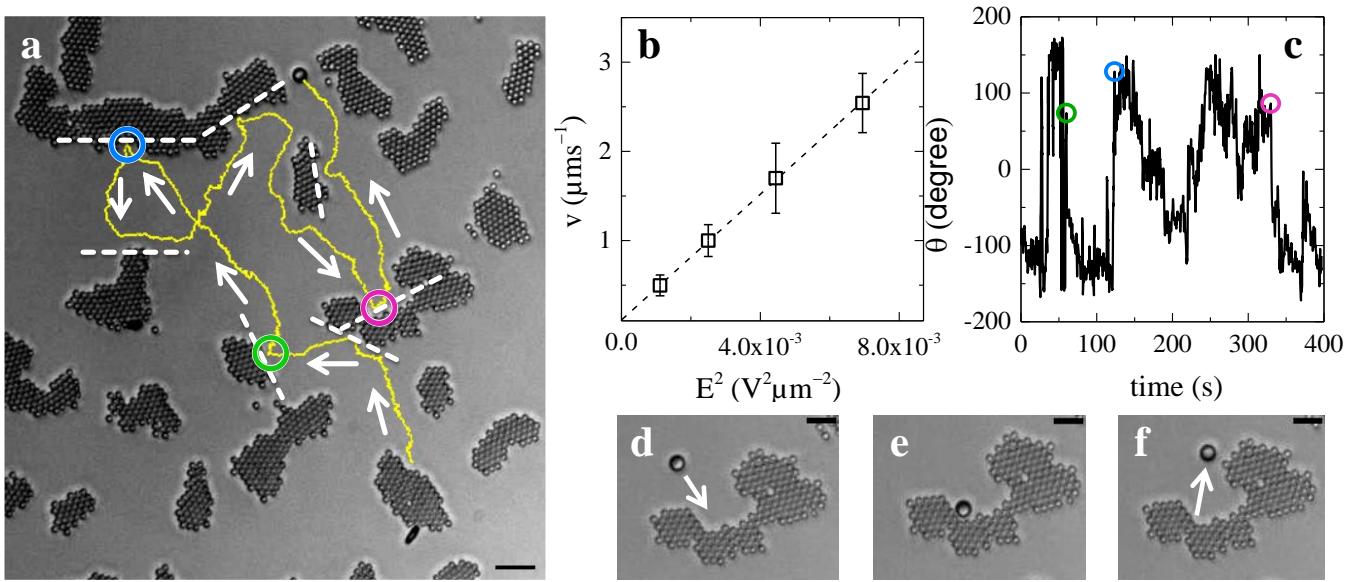

**Figure 3. Active JPs undergoing island-hopping in an environment of 'passive' colloids. (a)** Trajectory of an active JP (electric field, E= 0.066 V/μm) in the presence of passive silica spheres forming clusters (islands). The arrows and dashed lines indicate the swimming direction and the occurrence of "scattering events", respectively. The scale bar is 20 μm. **(b)** Mean swimming velocity, v, in the islands environment plotted against the squared electric field strength. Each data point is an average of over 10 – 15 individual particles based on the short-time MSD analysis. **(c)** Time evolution of the angle θ corresponding to the yellow trajectory in (a). The colored circles mark the same scattering events as in (a). **(d–f)** Sequence of snapshots showing the interaction of an active JP particle with an island of passive silica spheres; the microswimmer approaches and comes in contact with a cluster, reverses its orientation and swims away. The scale bar is 10 μm.

## 4 Discussion and Conclusions

Charged surfaces wetted by polar fluids and subjected to external electric fields develop slip flows due to the motion of the ions comprised in the electrical double layers. This leads to macroscopic fluid or particle transport[22,23], depending on whether the solid surface is fixed (as in the case of charged substrates) or freely suspended (as in the case of charged colloidal particles). A popular example is the migration of micro- and nanoparticles in the direction parallel or antiparallel to an applied DC field, E, with velocities proportional to the applied voltage (linear electrophoresis). Here, the island formation and active motion of JPs are driven by nonlinear electrophoretic effects known as induced-charge electroosmosis (ICEO) and induced-charge electrophoresis (ICEP).[24,25] The AC electric field perpendicular to the observation plane sets in motion the electrical double layer of the ITO-coated substrate and, simultaneously, polarizes the particle. The polarization changes the electric field just beneath the particle and drives the induced charges and solvent flow along the substrate.[26] The net result is the formation of recirculating electroosmotic rolls, where the current leaves and enters the surface and that can lead to an effective mutual attraction or repulsion between particles.[20,27] For silica colloids, the flow fields are attractive[19,28,29], consistently with our observation of close-packed islands at $f = 1.5$ kHz (**Fig. 1b**). In the case of silica-gold JPs, the metallic hemisphere is more strongly polarized by the applied electric field. This unbalance leads not only to local fluid rolls of different magnitude[20] but also to fluid being drawn from the front (ahead of the silica hemisphere) and ejected behind the gold cap.[18,21,30] Hence, self-propulsion occurs in the direction of the dielectric hemisphere.

Ultimately, the local fluid flows near silica and gold surfaces are, to our understanding, the reason behind the island-microswimmer 'scattering events' reported in **Fig. 3**. The JP is attracted by an island of 'passive' silica particles when it swims towards it with its $SiO_2$ hemisphere heading. As the microswimmer and the island are near to each other, the fluid rolls are strong enough to give rise to a hydrodynamic torque that quickly reorients the particle and, thus, the direction of its swimming velocity. Then, the Au hemisphere faces the island. In this situation the flow field of JP is opposite to that of the islands, and its repulsive flow field is sufficiently stronger than the inflow to provoke a quick detachment. The sequence is illustrated in **Supplementary Figure S2** and **Supplementary Video S2** where the island-microswimmer interaction is recorded at high-speed imaging. The outcome of these events is a JP-island collision that loosely resembles the phenomenon of 'elastic scattering', even though the Reynolds number of our macroparticles is always much smaller than 1 ($\mathrm{Re} \sim 10^{-6}$). Averaged over several collisions, the overall ABM of the JP has strong similarities with the swimming behavior of biological microorganisms undergoing run-and-tumble motion, with the islands acting as turning sites. The run-and-tumble dynamics manifests itself in the MSD of many JPs (**Figure 4**), showing a crossover from a ballistic to a diffusive regime at $\Delta t \sim d_i/v$, where $v$ is the mean swimming velocity of the microswimmers and $d_i$, is the mean inter-island distance (here $d_i$ plays the role of mean run length). Likewise, the tumbling rate is given by the inverse of the mean travel time between clusters. However, we note that the 'backscattering' does not show up in the MSD because the short-time behavior is dominated by the long ballistic runs.

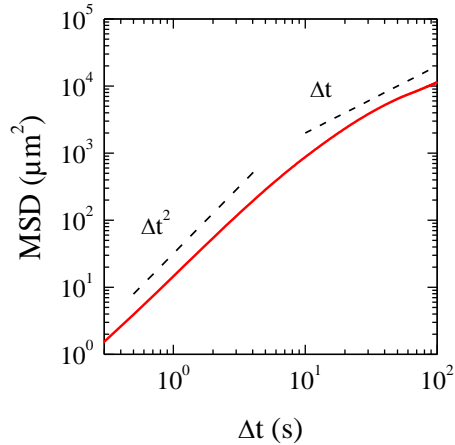

**Figure 4. ABM in the presence of island-hopping mechanism.** Mean squared displacement (MSD) of JPs at electric field 0.066 V/μm plotted on a double-logarithmic scale and averaged over 40 individual particle trajectories.

Overall, we demonstrated the dynamics of electric-field driven Janus particles in a bath of silica 'passive' spheres, which assemble into large clusters (islands) under the same electric field. Silica clusters have a significant influence on the active motion of the JPs. They act as turning sites by systematically reflecting the JP back upon collision. This backscattering mechanism is unusual in the absence of conventional inertia and is here as we propose due to the local electro-osmotic flows. Our work contributes to the rich library of dynamical effects displayed by 'passive' and 'active' colloidal particles subjected to AC electric fields. The reported motion type combines the intermittent propulsion of RTs and the long-time random walk of ABPs with a systematic reorientation mechanism. This work also poses a significant challenge to theoretical models, as the backscattering mechanism observed requires a deeper understanding of the interplay between the active and passive cluster.

## Acknowledgements

We thank Maheshwar Gopu, Laura Alvarez and Karthika Krishna Kumar for fruitful discussions. Financial support of the DFG (Grantno.PA459/18-2 and PA459/21-1) is gratefully acknowledged. We would like to acknowledge Michael Kappl and Leon Prädel from the Max Planck Institute for Polymer Research (MPIP), Mainz for their support in synthesis of the active colloids.

## Data availability statement
Original data are available from the authors upon reasonable request.

**Conflict of Interest**

The authors declare no conflict of interest.

**References**

(1) Stephen J. Ebbens; Jonathan R. Howse. In pursuit of propulsion at the nanoscale. *Soft Matter* **2010**, *6*, 726–738.

(2) Clemens Bechinger; Roberto Di Leonardo; Hartmut Löwen; Charles Reichhardt; Giorgio Volpe; Giovanni Volpe. Active particles in complex and crowded environments. *Rev. Mod. Phys.* **2016**, *88*, 45006.

(3) Aubret, A.; Ramananarivo, S.; Palacci, J. Eppur si muove, and yet it moves: Patchy (phoretic) swimmers. *Current Opinion in Colloid & Interface Science* **2017**, *30*, 81–89.

(4) Jie Zhang; Erik Luijten; Bartosz A. Grzybowski; Steve Granick. Active colloids with collective mobility status and research opportunities. *Chem. Soc. Rev.* **2017**, *46*, 5551–5569.

(5) Ivo Buttinoni; Giovanni Volpe; Felix Kümmel; Giorgio Volpe; Clemens Bechinger. Active Brownian motion tunable by light. *J. Phys.: Condens. Matter* **2012**, *24*, 284129.

(6) M. E. Cates; J. Tailleur. When are active Brownian particles and run-and-tumble particles equivalent? Consequences for motility-induced phase separation. *EPL* **2013**, *101*, 20010.

(7) Kilian Dietrich; Damian Renggli; Michele Zanini; Giovanni Volpe; Ivo Buttinoni; Lucio Isa. Two-dimensional nature of the active Brownian motion of catalytic microswimmers at solid and liquid interfaces. *New J. Phys.* **2017**, *19*, 65008.

(8) Juliane Simmchen; Jaideep Katuri; William E. Uspal; Mihail N. Popescu; Mykola Tasinkevych; Samuel Sánchez. Topographical pathways guide chemical microswimmers. *Nat Commun* **2016**, *7*, 1–9.

(9) Aidan T. Brown; Ioana D. Vladescu; Angela Dawson; Teun Vissers; Jana Schwarz-Linek; Juho S. Lintuvuori; Wilson C. K. Poon. Swimming in a crystal. *Soft Matter* **2015**, *12*, 131–140.

(10) Rodrigo Soto; Ramin Golestanian. Self-Assembly of Catalytically Active Colloidal Molecules: Tailoring Activity Through Surface Chemistry. *Phys. Rev. Lett.* **2014**, *112*, 68301.

(11) Juliane Simmchen; Paolo Malgaretti. Active Particle Accumulation at Boundaries: A Strategy to Measure Contact Angles. *ChemNanoMat* **2017**, *3*, 790–793.

(12) Lorenzo Caprini; Umberto Marini Bettolo Marconi. Active particles under confinement and effective force generation among surfaces. *Soft Matter* **2018**, *14*, 9044–9054.

(13) Claudio Maggi; Juliane Simmchen; Filippo Saglimbeni; Jaideep Katuri; Michele Dipalo; Francesco De Angelis; Samuel Sanchez; Roberto Di Leonardo. Self-Assembly of Micromachining Systems Powered by Janus Micromotors. *Small* **2016**, *12*, 446–451.

(14) Stenhammar, J.; Wittkowski, R.; Marenduzzo, D.; Cates, M. E. Light-induced self-assembly of active rectification devices. *Science Advances* **2016**, *2*, e1501850.

(15) Vutukuri, H. R.; Hoore, M.; Abaurrea-Velasco, C.; van Buren, L.; Dutto, A.; Auth, T.; Fedosov, D. A.; Gompper, G.; Vermant, J. Active particles induce large shape deformations in giant lipid vesicles. *Nature* **2020**, *586*, 52–56.

(16) Felix Kümmel; Parmida Shabestari; Celia Lozano; Giovanni Volpe; Clemens Bechinger. Formation, compression and surface melting of colloidal clusters by active particles. *Soft Matter* **2015**, *11*, 6187–6191.

(17) Joakim Stenhammar; Raphael Wittkowski; Davide Marenduzzo; Michael E. Cates. Activity-Induced Phase Separation and Self-Assembly in Mixtures of Active and Passive Particles. *Phys. Rev. Lett.* **2015**, *114*, 18301.

(18) Yan, J.; Han, M.; Zhang, J.; Xu, C.; Luijten, E.; Granick, S. Reconfiguring active particles by electrostatic imbalance. *Nature materials* **2016**, *15*.

(19) P. Richetti; J. Prost; P. Barois. Two-dimensional aggregation and crystallization of a colloidal suspension of latex spheres. *J. Phyique Lett.* **1984**, *45*, 1137–1143.

(20) Fuduo Ma; Xingfu Yang; Hui Zhao; Ning Wu. Inducing Propulsion of Colloidal Dimers by Breaking the Symmetry in Electrohydrodynamic Flow. *Phys. Rev. Lett.* **2015**, *115*, 208302.

(21) Sumit Gangwal; Olivier J. Cayre; MARTIN Z. BAZANT; Orlin D. Velev. Induced-Charge Electrophoresis of Metallodielectric Particles. *Phys. Rev. Lett.* **2008**, *100*, 58302.

(22) Stone, H. A.; Stroock, A. D.; Ajdari, A. Engineering Flows in Small Devices: Microfluidics Toward a Lab-on-a-Chip. *Annu. Rev. Fluid Mech.* **2004**, *36*, 381–411.

(23) Anderson, J. L. Colloid Transport by Interfacial Forces. *Annu. Rev. Fluid Mech.* **1989**, *21*, 61–99.

(24) Todd M. Squires; Martin Z. Bazant. Induced-charge electro-osmosis. *Journal of Fluid Mechanics* **2004**, *509*, 217–252.

(25) Martin Z. Bazant; Todd M. Squires. Induced-Charge Electrokinetic Phenomena: Theory and Microfluidic Applications. *Phys. Rev. Lett.* **2004**, *92*, 66101.

(26) W. D. Ristenpart; I. A. Aksay; D. A. Saville. Electrohydrodynamic flow around a colloidal particle near an electrode with an oscillating potential. *Journal of Fluid Mechanics* **2007**, *575*, 83–109.

(27) Songbo Ni; Emanuele Marini; Ivo Buttinoni; Heiko Wolf; Lucio Isa. Hybrid colloidal microswimmers through sequential capillary assembly. *Soft Matter* **2017**, *13*, 4252–4259.

(28) W. D. Ristenpart; I. A. Aksay; D. A. Saville. Electrically Guided Assembly of Planar Superlattices in Binary Colloidal Suspensions. *Phys. Rev. Lett.* **2003**, *90*, 128303.

(29) Trau, M.; Saville, D. A.; Aksay, I. A. Field-Induced Layering of Colloidal Crystals. *Science* **1996**, *272*, 706–709.

(30) Marjolein N. van der Linden; Lachlan C. Alexander; Dirk G. A. L. Aarts; Olivier Dauchot. Interrupted Motility Induced Phase Separation in Aligning Active Colloids. *Phys. Rev. Lett.* **2019**, *123*, 98001.

# Supplementary information
# for
# Island hopping of active colloids

Venkata Manikantha Sai Ganesh Tanuku[1]*, Peter Vogel[1], Thomas Palberg[1] and Ivo Buttinoni[2]

[1] Institute of Physics, Johannes Gutenberg University, 55128 Mainz, Germany
[2] Institute for Experimental Physics of Condensed Matter, Heinrich Heine University, 40225 Düsseldorf, Germany

## 1  Instantaneous velocity of active colloids

As shown in **Figure S1**, the calculated instantaneous velocity based on the JP trajectory in presence of silica clusters displayed in **Figure 3a** of the main text, reveals no significant change within the observed time interval. This also implicates that the particle velocity seems to be fully retained upon interaction with cluster islands as demonstrated for all events occurring within the 400 s time interval.

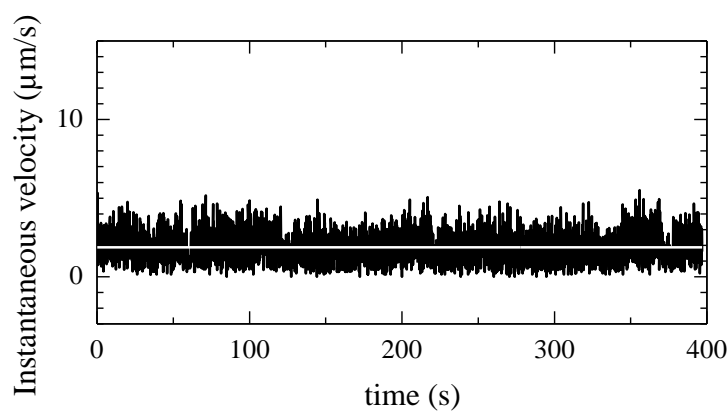

**Figure S1. Janus particle moving in a silica passive particle environment.** Instantaneous velocity calculated from the active particle trajectory in presence of silica cluster self-assembled in aqueous environments at constant electric field (E = 0.1 Vµm$^{-1}$). Within the observed time interval of $\Delta t$ = 400 s, the particle velocity fluctuates around its mean value (1.86 ± 0.02) µm/s represented by the horizontal line.

Obviously, we observed no pronounced drop in the instantaneous velocity during collisions of the active colloids and the particle velocity only fluctuates around the average.

## 2  Janus particle reorientation

To further support our interpretation of the JP – island interaction as scattering event accompanied by the reorientation of the gold cap to leave the cluster, we have performed high-speed imaging using the same system as shown in **Figure 3** of the main text. For the sake of clarity, we adjusted the brightness and contrast using ImageJ software to highlight and clearly show the gold cap orientation.

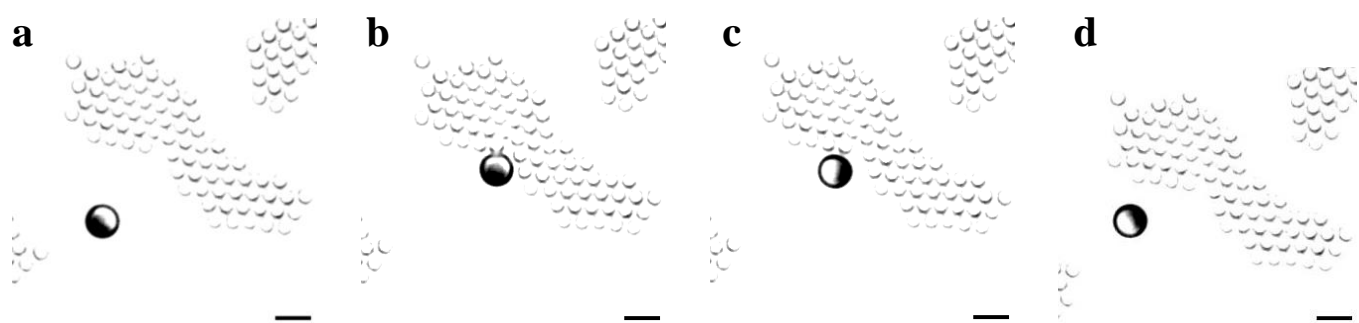

**Figure S2. High-speed imaging of the Janus particle – island cluster interaction.** Microscopic images (40 × magnification) taken in appropriate time intervals, $\Delta t$, at a high frame rate of 100 frames per second to reveal four distinct sequences for the interaction of silica-gold JPs with self-assembled silica clusters (E = 0.066 V/µm). **(a)** Displaying a representative JP localized near to the respective cluster with its silica side faced towards the island (starting point of analysis). The gold-coated side appears as dark hemisphere. **(b)** Landing of the JP (collision) while getting attracted by the silica cluster ($\Delta t$ = 2.2 s). **(c)** Gold cap reorientation ($\Delta t$ = 2.8 s). **(d)** Leaving the cluster through "back scattering" event. In each panel, the scale bar corresponds to 10 µm.

As demonstrated by the time series of microscopic images taken at high frame rates, the dynamic process of the microswimmer–island interaction occurs during considerably short time intervals (**Figure S2**). At the starting point of analysis shown in panel (a), the JP is attracted by an island of self-assembled smaller silica spheres while swimming with its uncoated silica side heading.

Upon getting in close contact with the cluster occurring around $\Delta t$ = 2.2 s ("collision event"), as displayed in panel (b), the particle starts rotating which results in a re-orientation of the gold coated hemisphere (dark side) shown in panel (c). The JP is quickly detached ("back scattering event") and starts leaving the cluster.

## 3   Supporting video information

To further illustrate the experimental findings regarding the island-formation after turning on the AC electric field, supporting video SV1 is recorded. Supplementary video SV2 shows the island-microswimmer interaction and the orientation of the gold-cap recorded at high-speed imaging.